\documentclass[preprint]{iucr}              

     \journalcode{S}              

\usepackage{comment}
\usepackage{amsmath}
\begin{document}                  



\title{A general method for multiresolutional analysis of mesoscale features in dark-field x-ray microscopy images}

\cauthor[a]{O.}{Abulshohoud}{shohomar@gmail.com}{}\aufn{On leave of absence from undergraduate study at the University of Chicago, \city{Chicago}, IL, \country{U.S.A}}
\author[a,b]{I.}{Poudyal}
\author[a]{J.}{McChesney}
\author[a]{Z.}{Zhang}
\author[a]{Z.}{Qiao}
\author[b]{U.}{Welp}
\cauthor[a]{Z.}{Islam}{zahir@anl.gov}

\aff[a]{X-Ray Science Division, Advanced Photon Source, Argonne National Laboratory \city{Lemont, IL}, \country{U.S.A}}
\aff[b]{Materials Science Division, Advanced Photon Source, Argonne National Laboratory, \city{Lemont, IL}, \country{U.S.A}}

\maketitle

\begin{synopsis}
Dark-field x-ray microscopy, which utilizes Bragg diffraction to collect full-field x-ray images on a crystalline sample, is a powerful tool being developed for determining the mesoscale structure of crystals. Information regarding the lattice structure and its physical implications is gleaned through the quantitative analyses of these images. Namely, one must be able to extract diffraction features that arise through contrast variations due to lattice heterogeneities, quantify said features, and track and identify patterns in the relevant quantitative properties in subsequent images. Due to the necessity to track features with a wide array of shapes and length scales while maintaining spatial resolution, we have employed wavelet transforms as a potent signal analysis tool. In addition to addressing multiple length scales, this method can be used in conjunction with other signal processing methods such as image binarization for increased functionality. 
\end{synopsis}

\begin{abstract}
    Dark-field x-ray microscopy utilizes Bragg diffraction to collect full-field x-ray images of 'mesoscale' structure of ordered materials. Information regarding the  structural heterogeneities and their physical implications is gleaned through the quantitative analyses of these images. Namely, one must be able to extract diffraction features that arise from lattice modulations or inhomogeneities, quantify said features, and identify and track patterns in the relevant quantitative properties in subsequent images. Due to the necessity to track features with a wide array of shapes and length scales while maintaining spatial resolution, wavelet transforms was chosen as a potent signal analysis tool. In addition to addressing multiple length scales, this method can be used in conjunction with other signal processing methods such as image binarization for increased functionality. In this article, we demonstrate three effective use of wavelet analyses pertraining to DFXM. We show how to extract and track smooth linear features\textemdash which are diffraction manifestations of twin boundaries\textemdash as the sample orientation changes as it is rotated about momentum transfer. Secondly, we show that even the simplest wavelet transform, the Haar transform, can be used to capture the primary features in DFXM images, over a range of length scales in different regions of interest within a single image enabling localized reconstruction. As a final application, we extend these techniques to determine when a DFXM image is in focus 
\end{abstract}

\section{Introduction} 
Mesoscale defects~\cite{crabtree2012quanta} in a lattice structure greatly affect the physical properties of the associated crystal, such as its tensile strength, electromagnetic behavior, and emergent properties in neuromorphic and quantum materials \cite{simons2018long}. Therefore, it is imperative to identify and understand how defects transform when a material undergoes various physical effects such as thermal cycling and applied fields.
\par In this work, we are concerned with the images gathered using dark-field x-ray microscopy (DFXM), which utilizes Bragg diffraction to identify strain, misorientation, and other local heterogeneities \cite{simons2015dark,poulsen2018reciprocal,yildirim2020probing,bucsek2019sub,dresselhaus2021situ}. By analyzing the contrast variations in the collected image that arise from these disparities, local variations in the lattice configuration can be discovered. 
\par There are several possible approaches for acquiring information about crystalline structures from dark-field x-ray images, all of which require quantitative extraction, classification, and tracking of features within a collection of two-dimensional (2D) diffraction images~\cite{brennan2022ana}. The two main approaches are to collect projection images
from a single perspective as a function of time~\cite{poudyal2022pump} or external perturbations~\cite{dresselhaus2021situ}, or to collect a series of projections~\cite{ludwig2001three} from different perspectives and reconstruct a three-dimensional (3D) image~\cite{flannery1987three}. \par
A collection of diffraction images representing stand-alone two-dimensional projections of the crystal as it is subject to external stimuli\textemdash such as a change in temperature, application of strain~\cite{dupraz20173d}, or being subject to an electromagnetic field~\cite{karpov2017three}\textemdash is alone a powerful method for characterizing a material. As the defects in the lattices transformed, it would be necessary to quantitatively analyze these features to understand the physical implications of their motion and compare the results to the theoretical expectation. If necessary, one could reconstruct this collection of projections into a full three-dimensional model of the defect network in the sample, from which two-dimensional ``slices" of interest can be isolated to study separately. Observing these slices again requires an algorithm to quantify the structure and motion of important features ({\it e.g.} dislocations, magnetic or ferroelectric domain walls, charge density wave phases).
\par In any of the general strategies discussed above, one invariably requires the quantitative extraction and tracking of features with elevated intensities in diffraction images. In this paper, we develop a methodology of performing said analysis. We first codify the required breadth and depth of the methodology through observation of data and domain-science knowledge. We then explore potential signal analysis tools that would enable the methodology to complete the analyses most accurately and efficiently. We ultimately develop these techniques into a full algorithm, and we apply this algorithm to a set of DFXM images, in order to analyze performance and determine the next steps to be taken. 
\section{Experimental Technique}
In order to convert the collected DFXM image analysis data into meaningful scientific information, it is necessary to apply the algorithms with an understanding of the crystallographic techniques used (e.g. an understanding of how particular crystal defects would manifest in diffraction images). The details of DFXM\textemdash the motivating experimental technique of this article\textemdash and the images we are concerned with analyzing are discussed in this section.

\subsection{Dark-field X-ray Microscopy}
\begin{figure}
    \centering
    \caption{A generic schematic of DFXM setup. An x-ray beam of a fixed wavelength illuminates a crystalline sample. A diffracted beam then passes through an objective lens to form a magnified image on a detector.} 
    \includegraphics[angle=0,width=1.0\textwidth]{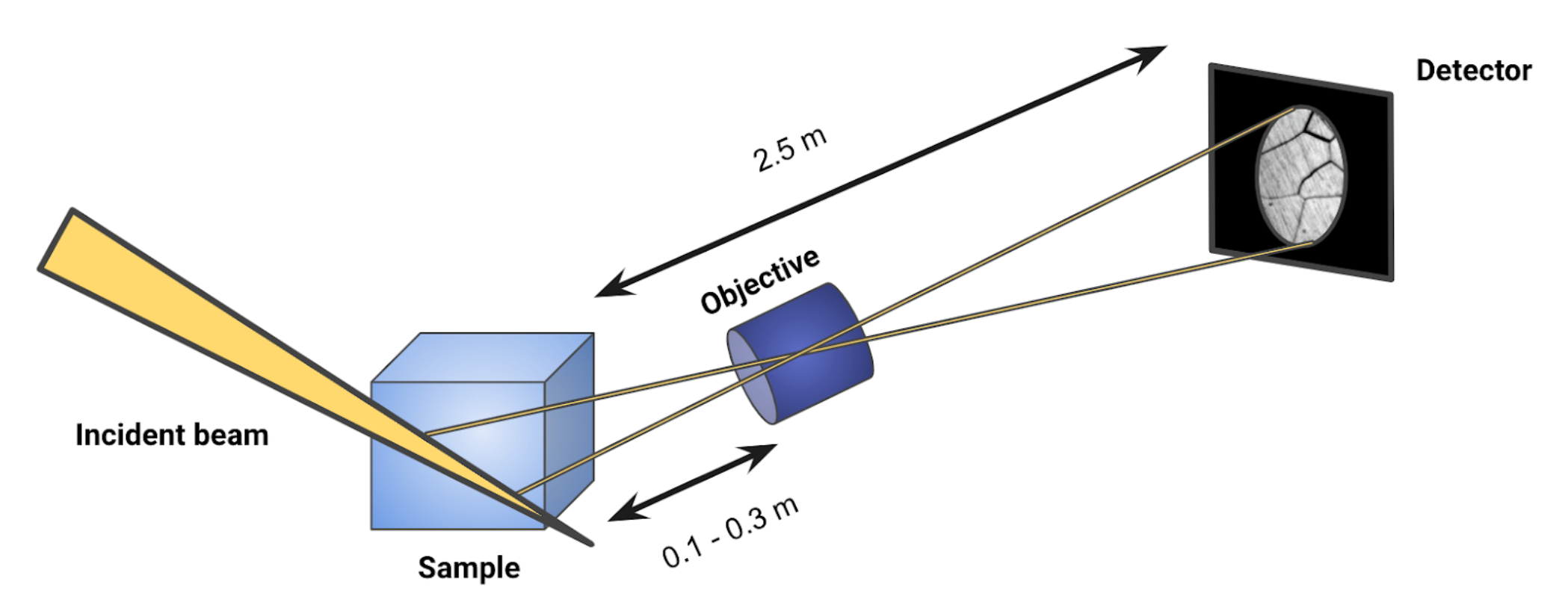}
    \label{fig:dfxm}
\end{figure}
 \par DFXM is a full-field imaging technique which creates real-space images on mesoscale structures in a sample of ordered material~\cite{simons2015dark,poulsen2018reciprocal}. Observing contrast variations caused by local disparities in the lattice structure enables the capturing of various physical properties (such as phase, orientation, and strain) of crystal deformities that are deep below the sample surface which can be directly probed by hard x-rays. Detectors are typically placed several meters away (in our case, 2.5 m) from the sample, with objective-lens focal length long enough to  provide sufficient working distance for the use of large sample environments \textit{in situ}.
 \par The inclusion of an x-ray objective lens in the diffracted beam, as shown in Figure \ref{fig:dfxm}, is an essential feature of DFXM. By adjusting the focal length of the lens, the user can zoom in and out to investigate a wide range of length scales, thus capturing a broad array of mesoscale features while maintaining a precise spatial as well as angular resolution. The objective lens also acts as a filter for stray diffracted beams, thereby isolating elements of interest within the sample.\cite{yildirim2020probing}

\subsection{Experimental Details}
The DFXM images in Figures \ref{fig:imageset}A and \ref{fig:imageset}B were collected on 33-ID-D at the Advanced Photon Source using 10 keV x-rays with a bandwidth of 1.3 eV. In this case a KB mirror pair was used as the condenser lens and an x-ray zone plate as the objective lens.  An Andor Neo sCMOS camera with a pixel size of 6.5 $\mu$m $\times$ 6.5 $\mu$m was used as the detector in combination with a thin-film scintillator and a 20x optical lens, yielding a total magnification of 400x. The images were captured with a viewing angle of $80^\circ$ about the horizontal axis from surface normal, yielding an effective pixel size of 16 nm $\times$ 33 nm. 
\par The DFXM image in Figure \ref{fig:imageset}C was collected on 6-ID-C at the Advanced Photon Source using 13 and 20 keV x-rays, repectively. The microscope utilized either a polymeric (20 keV) or Be (13 keV) refractive lens as a condenser, and a polymeric (20 keV) refractive lens as an objective, without any harmonic rejection mirrors. An Andor 5.5MP Zyla sCMOS camera was used with a scintillator followed by a 5X optical objective lens, yielding a total magnification of $\sim$80X~\cite{qiao2020large}. 

\subsection{DFXM Images}
\begin{figure}
    \centering
	\includegraphics[angle=0,width=1.0\textwidth]{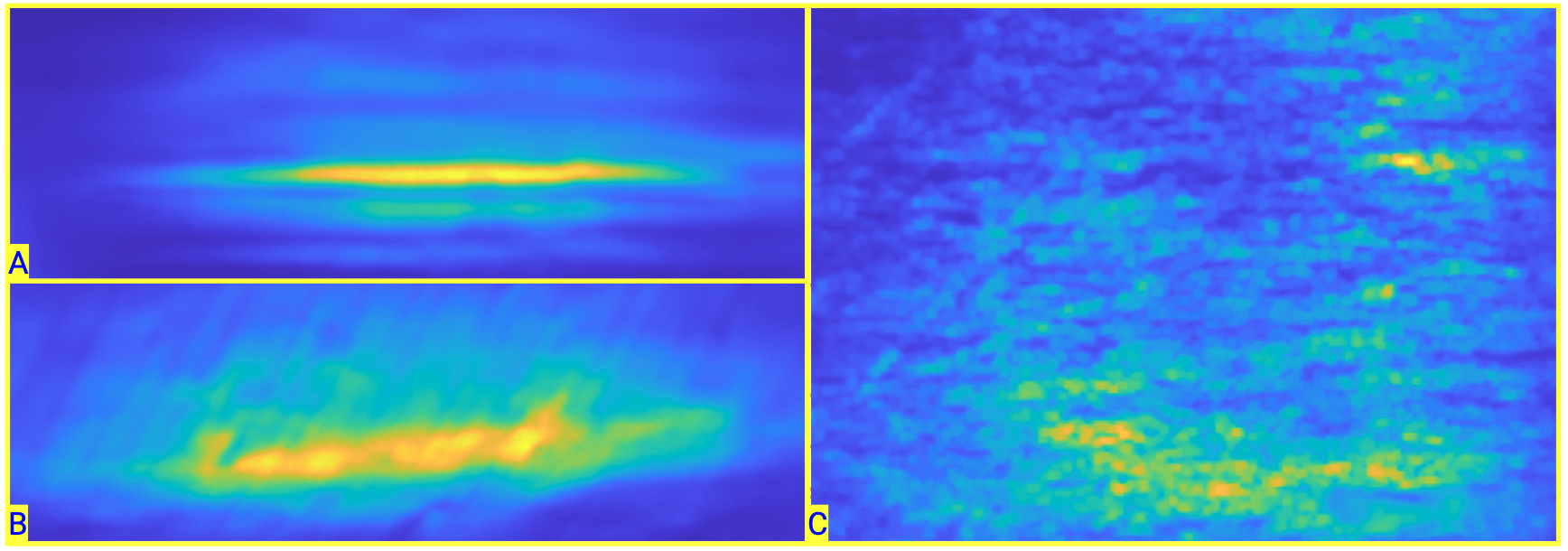}
    \caption{A collection of various DFXM images from single-crystal samples. Images A and B contain a perspective distortion caused by a viewing angle of $80^\circ$ about the horizontal axis from surface normal. \\
    A: DFXM image of twin boundaries in yttrium barium copper oxide (YBCO) crystal.\\
    B: DFXM image of twin boundaries in YBCO after an $84^\circ$ rotation of the sample relative to Figure A. \\
    C: DFXM image of single crystal strontium titanate (STO).  \\}
    \label{fig:imageset}
\end{figure}

\par
\begin{figure}
    \centering
	\includegraphics[angle=0,width=1.0\textwidth]{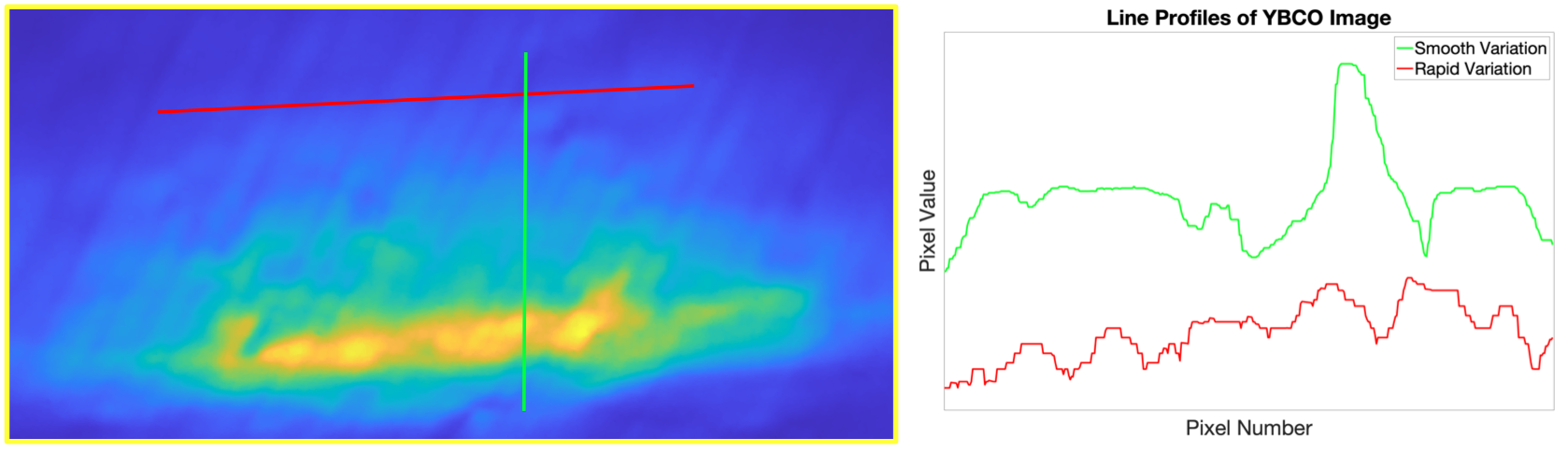}
    \caption{Near-orthogonal line profiles of Figure \ref{fig:imageset}B. The line profile between boundaries (red) captures more rapid oscillations, whereas the changes depicted by the line profile along an image column (green) shows smoother variations. This demonstrates the need for a versatile signal analysis tool, one which can extract large- and small-periodicity features.}
    \label{fig:doublelp}
\end{figure}

 The features captured in relevant DFXM images take on a diverse array of shapes, sizes, and regularity\textemdash contrast variations range from linear features with an underlying smooth variation (Figures \ref{fig:imageset}A and \ref{fig:imageset}B), to point-like details \cite{gonzalez2020methods}, to features with nondescript shapes \cite{bucsek2019sub}, to a combination of the three at varying length scales (Figure \ref{fig:imageset}C), all of which represent different configurations of irregularities and their orientations. Thus, in order to function properly for the multitude of possible defects and their network , the algorithm used to analyze DFXM images must be versatile with respect to the sizes, shapes, locations, and frequency of features. Furthermore, to identify the boundaries of both arbitrary gross features and the details contained within them, the signal analysis tool used must be able to capture smooth and rapid variations of up to three orders of magnitude apart, encompassing the mesoscale. For example, as depicted in Figure \ref{fig:doublelp}, the tool must capture both the smooth variations along the twin boundaries and the rapid oscillations between them in order to separate features of interest from the overall profile of the illuminating X-rays. It also must be capable of extracting features and oscillations of varying shapes, and must be localized in space so that the position of said features can be resolved. These requirements are satisfied by wavelets and the wavelet transform~\cite{walker2008primer}.

\section{Image Processing}
\par While the wavelet transform will be the primary technique featured in this paper, other tools to be used in conjunction with it are also considered, specifically image binarization. This section discusses the nature and application of said tools. As the primary signal analysis technique used in this project, the applications of the wavelet transform were explored first.

\subsection{Discrete Wavelet Transform}
\par The wavelet transform is a method of expressing a signal in a basis of “wavelet” functions, which are localized in space and periodicity. The wavelets can be scaled and translated to provide multi-resolutional analysis \cite{heil1989continuous}. In contrast, Fourier Analysis would give only periodicity without space localization. While this may be sufficient for images that contain single features, it would not be ideal when desiring to extract features of different length scales in different regions of an image.  Furthermore, the sinusoidal restriction of Fourier analysis leaves less versatility when compared to the myriad of possible wavelets available with the wavelet transform, which complicates analyses for features than may be particularly well-extracted by a matching wavelet. \par
Applying a wavelet transform to an image yields two sub-images\textemdash one is an approximation of the original image, identifying smooth variations, while the other is a detail image, which identifies rapid oscillations or abrupt changes in features. The wavelet transform can be repeated on the approximation signal to yield yet another set of sub-images.
In developing this methodology, we sought to first develop an intuition with the simplest images and wavelets and utilize it to optimize the algorithm, before moving on to more complex data \cite{shensa1992discrete}. Accordingly, we first analyzed Figure \ref{fig:imageset}A with the Haar wavelet.
\subsubsection{The Haar Transform}
\par Having only two values, the Haar wavelet is generally thought to be the simplest wavelet. Furthermore, due to its shape (essentially a single period of a square wave), the Haar wavelet excels at detecting and storing sharp edges within features. Therefore, the horizontal boundaries with clear distinctions present in Figure \ref{fig:imageset}A would hypothetically be aptly captured by the Haar transform. 
\par The Haar wavelet is one of few wavelets with simple and intuitive associated transforms. Adjacent data in a signal are connected in a line. The midpoint of the line is stored in the approximation signal, and the slope of the line (scaled by a factor of $-\frac{1}{2}$) is stored in the detail signal. Both subsignals are then scaled by an additional factor of $\sqrt{2}$, to preserve the total energy of the signals. Further details on the Haar transform are discussed in \citeasnoun{walker2008primer}.
\begin{figure}
	\centering
	\includegraphics[angle=0,width=1\textwidth]{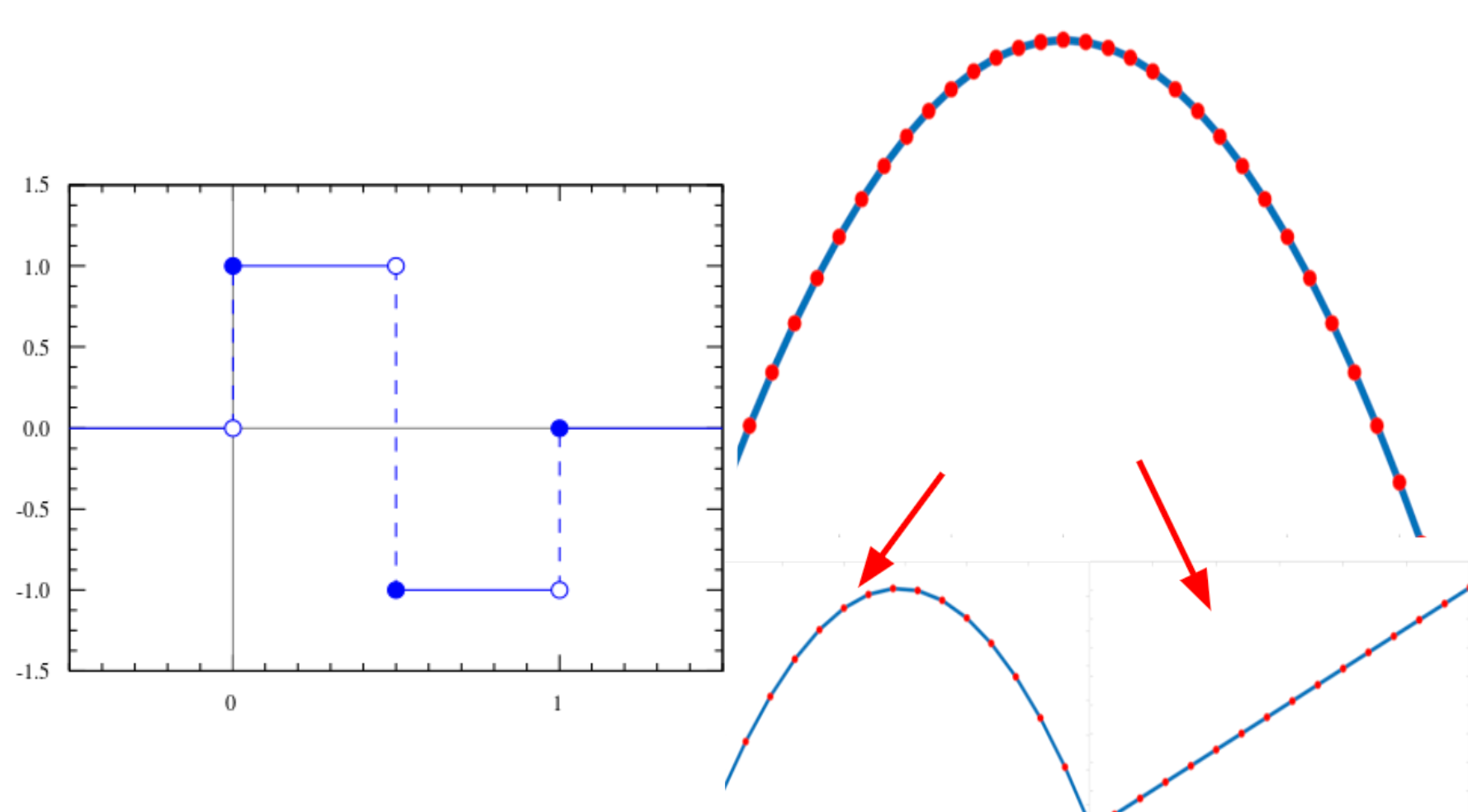}
    \caption{Left: A graph of the Haar wavelet in the space domain. Right: An example of a Haar transform on an inverted parabola, with approximation and detail signals.}
    \label{fig:haarwav}
\end{figure}
\par

\subsubsection{Selective Reconstruction}
\par
\begin{figure}
	\centering
	\includegraphics[angle=0,width=1\textwidth]{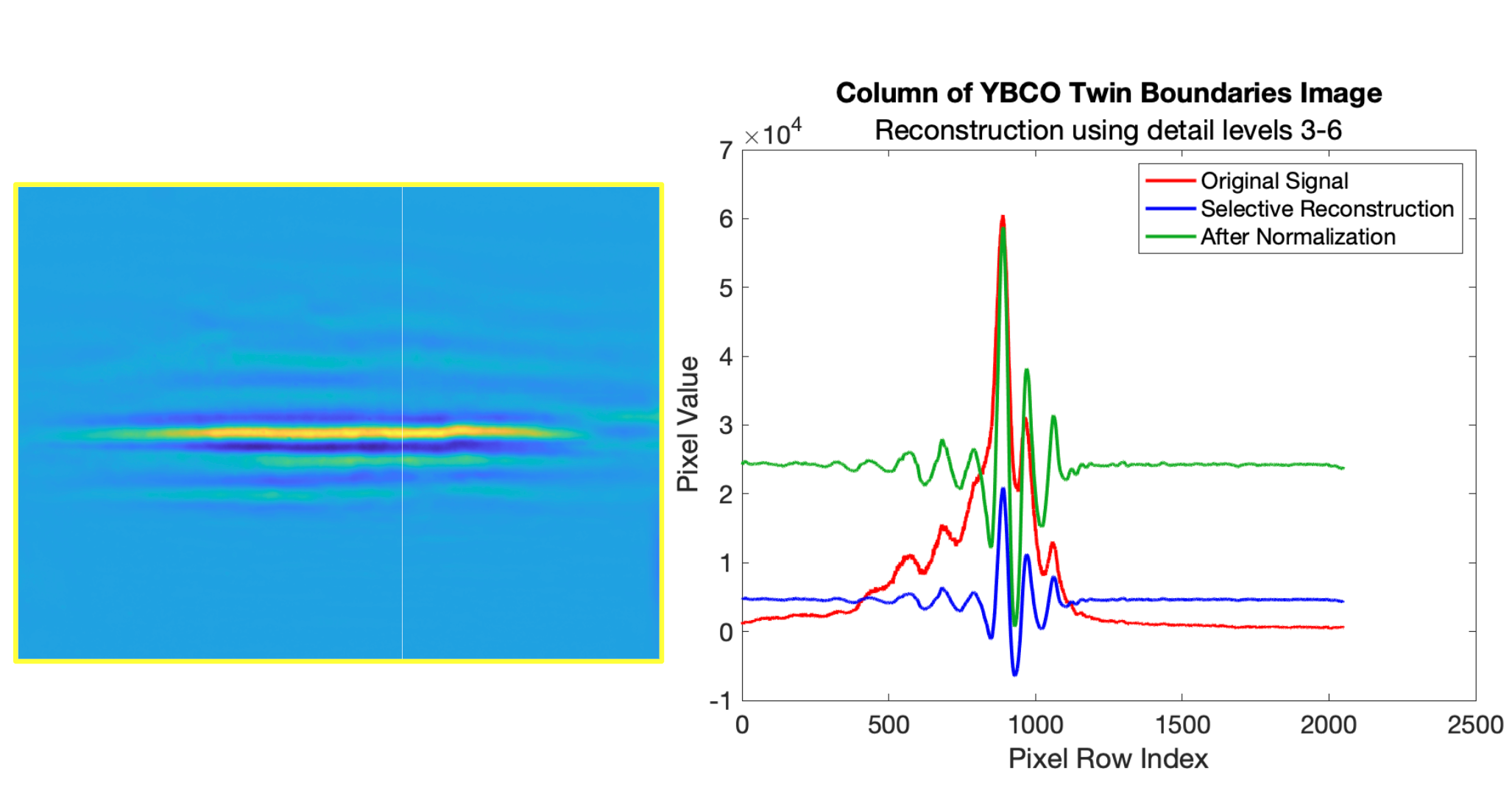}
    \caption{Left: YBCO image twin boundaries image (Figure \ref{fig:imageset}A) after selective reconstruction and renormalization is applied, with a column of interest highlighted. Right: Plot of original signal, selective reconstruction, and normalized selective reconstruction of the column. Selective reconstruction removes the near Gaussian profile, resulting in the features of interest oscillating about the same mean value.}
    \label{fig:sr}
\end{figure}

 In the process of reconstructing an image using the inverse wavelet transform on sub-images, undesired detail signals can be omitted by being set to zero. This is beneficial because it allows the isolation or omission of features of particular lengths or periodicities. For example, in Figure \ref{fig:imageset}A, there is a near Gaussian envelope over the image signal, causing central features to be brighter than those at the edges, which is likely the results of the illuminating X-ray profile. Furthermore, the camera is subject to small-periodicity noise across the entire image (caused by experimental imperfections in background light omission) which complicates the quantitative extraction of gross features and their properties. Utilizing selective reconstruction allows the removal of the smooth near Gaussian profile (i.e. normalizing out the illumination function) and the rapid noise oscillations, preserving the mid-periodicity features of importance, as shown in Figure \ref{fig:sr}. Upon examining the line profiles of the Haar approximations and details at different levels\textemdash with detail level 1 representing changes on the smallest length scale and detail level 11 representing the largest\textemdash levels 7-11 were shown to encapsulate the details of the broad near Gaussian envelope, while levels 1-2 contained high-frequency noise that did not significantly change the contour of the image. Thus, levels 3-6 were determined to be the ideal detail levels for capturing the mid-frequency features of interest. More details regarding the specific mathematics of selective reconstruction, as well as the analysis of appropriate detail levels, are discussed in Appendix \ref{sec:appendixsr}. 

The wavelet transform excels at bringing the diffraction features of interest into focus. However, to see how this way of encapsulating features is beneficial, we must determine our quantitative extraction method. This brings us to image binarization.
\subsection{Image Binarization}
\par While the wavelet transform is powerfully suited to the requirements of diffraction image analysis, it is not the only tool available. One of the most common techniques utilized for extracting quantitative properties of features with elevated intensity is image binarization, which can be used in conjunction with the wavelet transform to yield more useful results.
\par Binarizing an image is the act of converting each pixel into a boolean variable, based on whether the value of the pixel satisfies a given condition. Typically, the condition used to determine binarization is whether a pixel is above or below a certain threshold. Pixels above the threshold are set to ``on", while pixels below are set to ``off". Thus, binarization rigorously defines previously ambiguous shapes and boundaries, thus enabling them to be quantitatively analyzed. Properties of regions of interest that can be measured include area, centroid, diameter, and many other geometric quantities. The algorithm for binarization was adopted from \citeasnoun{tamraz}. 
\par
\begin{figure}
	\centering
	\includegraphics[angle=0,width=1\textwidth]{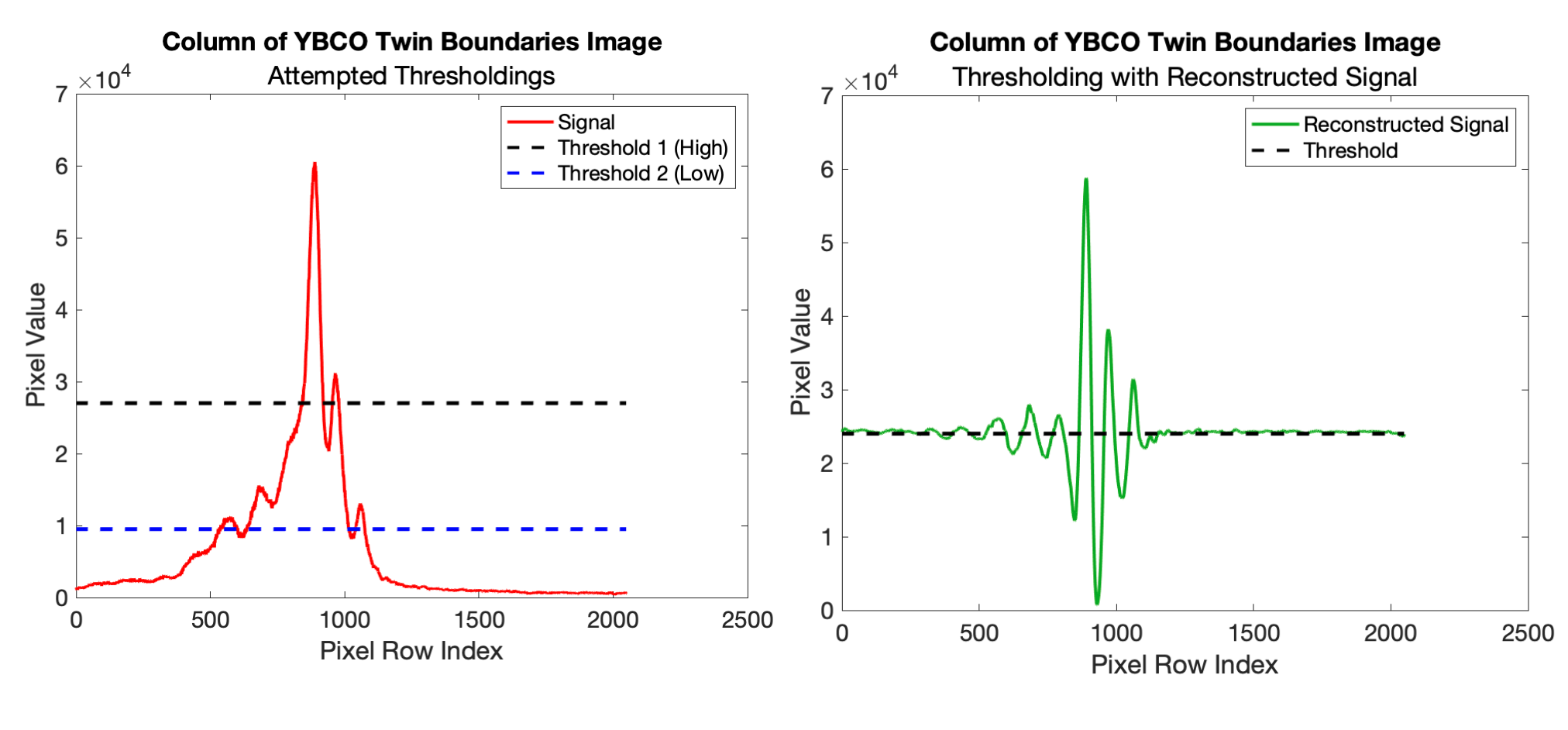}
    \caption{Left: Profile of column of interest in original signal with attempted thresholds. Right: Profile of column of interest in selectively reconstructed signal with potential threshold. Applying selective reconstruction before binarizing allows low-intensity features to be captured without sacrificing distinctions between high-energy features.}
    \label{fig:srbin}
\end{figure}

 As indicated in Figure \ref{fig:srbin}, when attempting to binarize the original image, distinctions between high-intensity features are lost if low-intensity features are captured, and vice versa. However, upon selectively reconstructing the image, the features of interest are made to oscillate about a common mean, enabling all of them to be captured without losing distinctions. Thus, applying selective reconstruction before binarization improves the accuracy of the feature extraction. Ultimately, while binarization is used to quantify the features of interest, the chosen wavelet transform is what truly captures them and brings them into focus.

\section{Discussion}
The image analysis algorithm can be separated into two broad sections\textemdash extracting features within an image, and tracking how features change between contiguous images. In this section, results for feature extraction are discussed in-depth, while an outline of a potential feature tracking algorithm is given.

\subsection{Feature Extraction}
\par
\begin{figure}
    \centering
	\includegraphics[angle=0,width=0.9\textwidth]{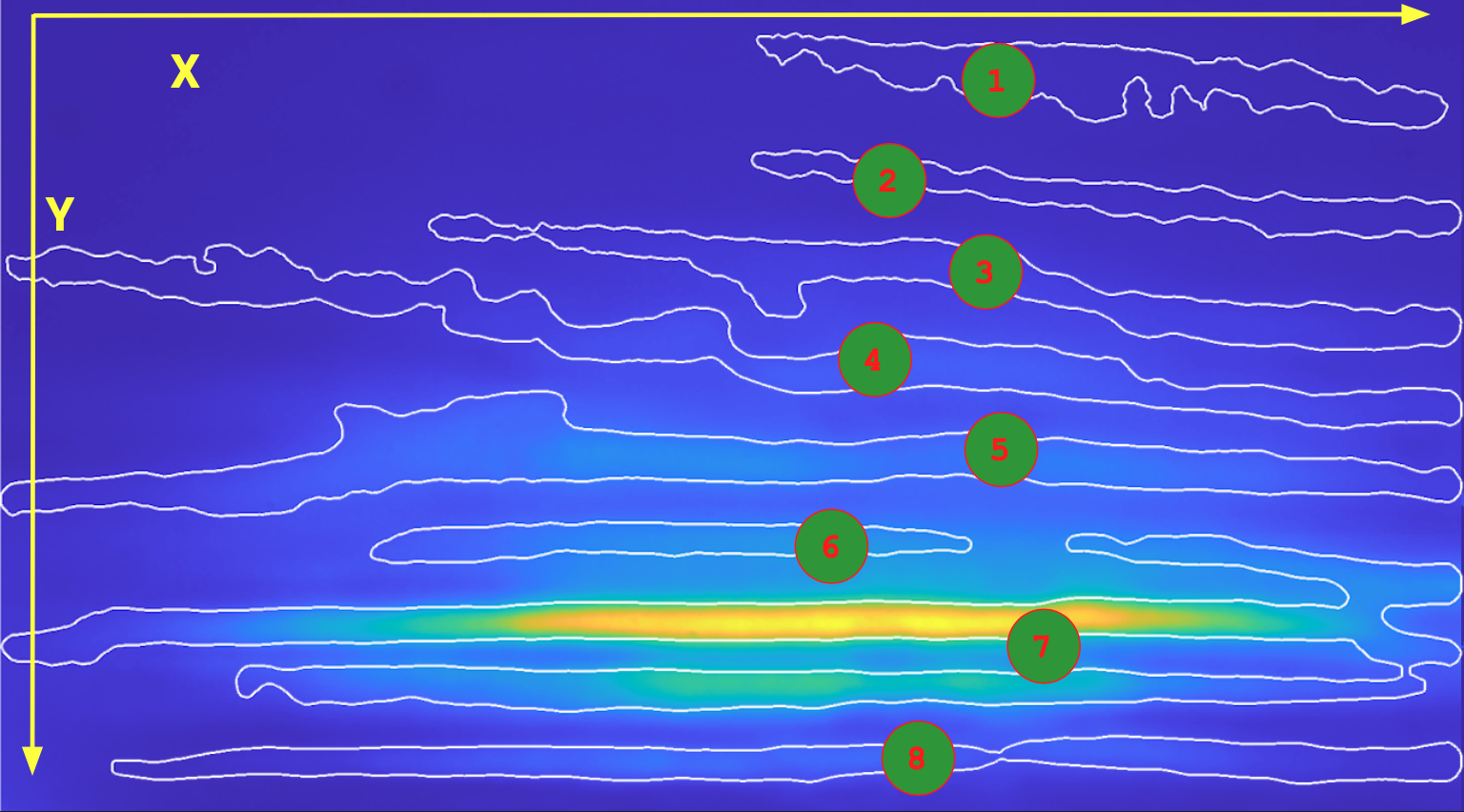}
    \caption{Results of the binarization using selective reconstruction, with the binary region outlines overlaid on the original image. Most linear features of note are successfully encapsulated by the outlines in a way that stays true to their orientation, shape and size.}
    \label{fig:extract}
\end{figure}

 Selective reconstruction with the Haar transform was used to denoise the image and separate the low-frequency near Gaussian envelope. The feature clustering algorithm was then used to remove the exterior background of the image. The image was then slightly refined using a Gaussian filter to smooth out the peaks and was finally extracted using binarization. Before feature quantification, a final refinement was implemented by removing low-area features. The results are shown in Figure \ref{fig:extract}. The method successfully outlines gross features within the image and is able to extract important quantitative properties, as shown in Table \ref{tab:extractq}.
\par
\begin{figure}
	\centering
	\includegraphics[angle=0,width=1\textwidth]{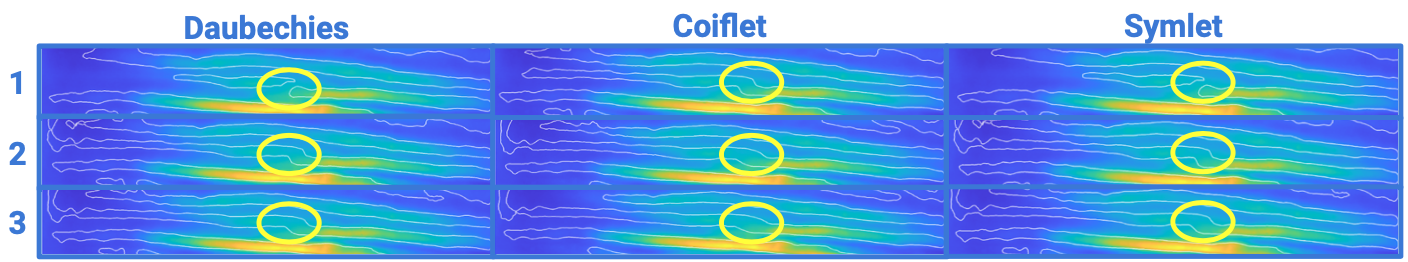}
    \caption{Extraction results of randomly-selected YBCO image after selective reconstruction with Daubechies wavelets 1-3, coiflets 1-3, and symlets 1-3.}
    \label{fig:wavecol}
\end{figure}

 The majority of notable linear features within the image were successfully and accurately captured by the algorithm. The most significant error is in feature 7, which absorbed the tail of feature 6 and was connected on the right side of the image with a narrow bridge to what is seemingly a different feature but was not extracted separately. These errors may be mitigated or resolved with the use of a more intricate wavelet.
\begin{table}
\resizebox{\textwidth}{!}{%
\begin{tabular}{|p{0.09\textwidth}|p{0.075\textwidth}|p{0.13\textwidth}|p{0.1\textwidth}|p{0.13\textwidth}|p{0.15\textwidth}|p{0.1\textwidth}|p{0.12\textwidth}|}
\hline
Feature Number & Area ($\mu$m$^2$) & Centroid ($\mu$m) & Diameter ($\mu$m) & Eccentricity & Orientation (degrees from $x$-axis) & Average Intensity (0-1) & Integrated Intensity (pixels) \\
    \hline
        1 & 70.3 & (22.8, 9.44) & 15.8 & 0.996 & -25.2 & 0.0307 & 1420 \\ \hline
        2 & 49.5 & (23.4, 23.6) & 16.4 & 0.999 & -27.8 & 0.0397 & 1290 \\ \hline
        3 & 91.0 & (20.3, 34.0) & 25.4 & 0.998 & -33.5 & 0.0804 & 4820 \\ \hline
        4 & 137 & (14.9, 40.5) & 35.7 & 0.999 & -33.1 & 0.104 & 9340 \\ \hline
        5 & 166 & (14.2, 55.0) & 30.0 & 0.998 & 0.348 & 0.191 & 20900 \\ \hline
        6 & 42.2 & (13.5, 64.5) & 12.4 & 0.998 & 2.96 & 0.265 & 7350 \\ \hline
        7 & 249 & (16.8, 76.0) & 30.4 & 0.995 & 8.18 & 0.431 & 70500 \\ \hline
        8 & 96.7 & (16.9, 90.4) & 27.7 & 1.00 & 1.13 & 0.126 & 8000 \\ \hline
    \end{tabular}
 }
	\caption{Quantitative properties of extracted features, corrected for the perspective distortion caused by the viewing angle of $80^\circ$ from surface normal. Each of these properties can be utilized to determine the size, shape, location, orientation, and brightness of the feature. Tracking features across frames will allow the investigation of how these features change over time.}
	\label{tab:extractq}
\end{table}
Upon developing the methodology with the simple and intuitive Haar wavelet, other wavelet transforms were tested to compare performance. Figure \ref{fig:wavecol} shows the extraction results in a region of interest of a randomly-selected image of the YBCO sample undergoing rotation for nine different wavelets: Daubechies 1 through 3, coiflet 1 through 3, and symlet 1 through 3. The Daubechies 1 wavelet is equivalent to the Haar wavelet. 

The central feature in the images undergoes significant variance in the extraction, with certain wavelets mistakenly extracting it as two separate features. Ultimately, Daubechies 2 was thought to provide sufficient extraction potency without significant computational demand\textemdash all subsequent analyses are done with the Daubechies 2 wavelet transform.
\subsection{Localization of Reconstruction}
\begin{figure}
	\centering
	\includegraphics[angle=0,width=1.0\textwidth]{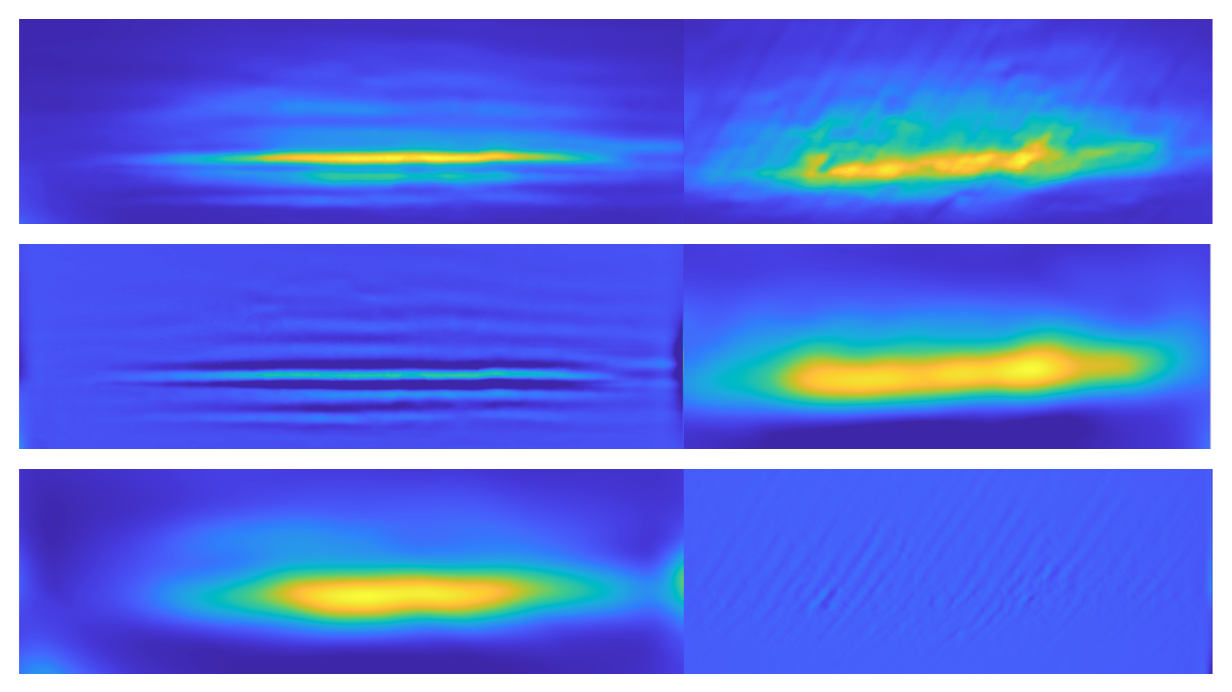}
    \caption{Top: Concatenation of Figure \ref{fig:imageset}A and \ref{fig:imageset}B. Middle: Selective reconstruction of concatenated image, with rapid-variation reconstruction on the left half and smooth-variation reconstruction on the right. Bottom: Selective reconstruction of concatenated image, with smooth-variation reconstruction on the left half and rapid-variation reconstruction on the right.}
    \label{fig:locdep}
\end{figure}

The spatial resolution of the wavelet transform allows distinct regions within images to be selectively reconstructed using different detail levels. This provides the ability to analyze different length scales in separate parts of the image, to accord with the sizes of features of interest in different regions. To demonstrate this capability, Figures \ref{fig:imageset}A and \ref{fig:imageset}B were concatenated into a single image. This concatenation was then locally reconstructed with location-dependent detail levels, and the results are shown in Figure \ref{fig:locdep}. The central concatenated image in the figure shows a small-length-scale reconstruction of Figure \ref{fig:imageset}A and a larger-length-scale reconstruction of \ref{fig:imageset}B, while the bottom image shows the inverse.

\subsection{Feature Tracking}
\label{sec:track}
\par
\begin{figure}
	\centering
	\includegraphics[angle=0,width=0.9\textwidth]{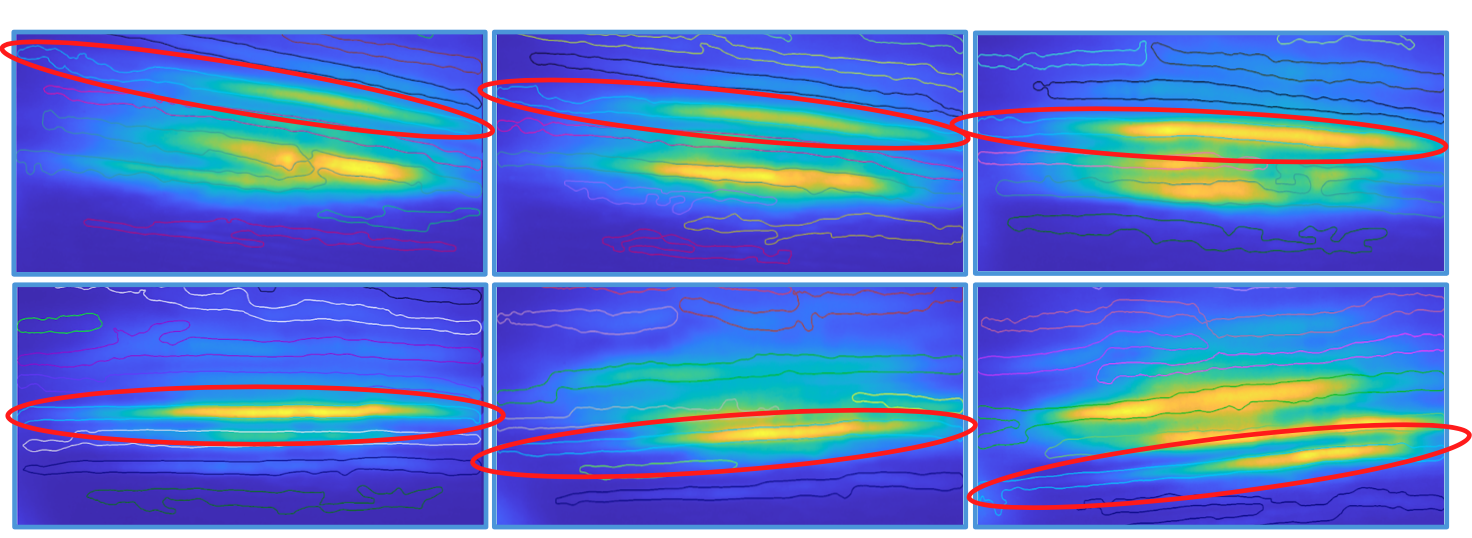}
    \caption{DFXM images on YBCO twins through a $45^\circ$ sweep ($9^\circ$ increments) with multicolored outlines to show tracking algorithm results. The red ovals encapsulate a contrast feature that the tracking algorithm correctly identifies to be produced by the same real-space mesoscale heterogeneity in each image, as indicated by the consistent blue highlight around the feature.}
    \label{fig:tracker}
\end{figure}

 Upon subjecting a sample of ordered material to physical perturbations, images may be taken in quick succession to construct a DFXM video showing how the Bragg signature of features changes over time. In this case, features only undergo minor changes from image-to-image, and thus an extracted feature may be tracked throughout the process to observe how its physical properties change over the course of their ``lifetime" of visibility. Thus, an algorithm to track features was developed. Upon extracting features in adjacent images, the algorithm computes their ``feature distance", a measure of dissimilarity of their quantitative properties. Each feature is then paired with the feature most similar to it in the opposite image, provided that their feature distance falls below a maximum threshold. Unpaired features are deemed to have newly disappeared or appeared. Diffraction features between images that the algorithm claims to be linked are outlined in the same color. The results of this tracking algorithm implemented through a 45$^\circ$ rotation of the YBCO sample are shown in Figure \ref{fig:tracker}. 

The algorithm successfully tracked a feature throughout the  sweep, and several others were successfully tracked throughout broad increments. The algorithm misidentifies some features as new despite their clear relation to previously extracted features, which is likely due to an erroneously low feature distance cut-off between paired features.

\subsection{Finding Image Focus}
\begin{figure}
	\centering
	\includegraphics[angle=0,width=1\textwidth]{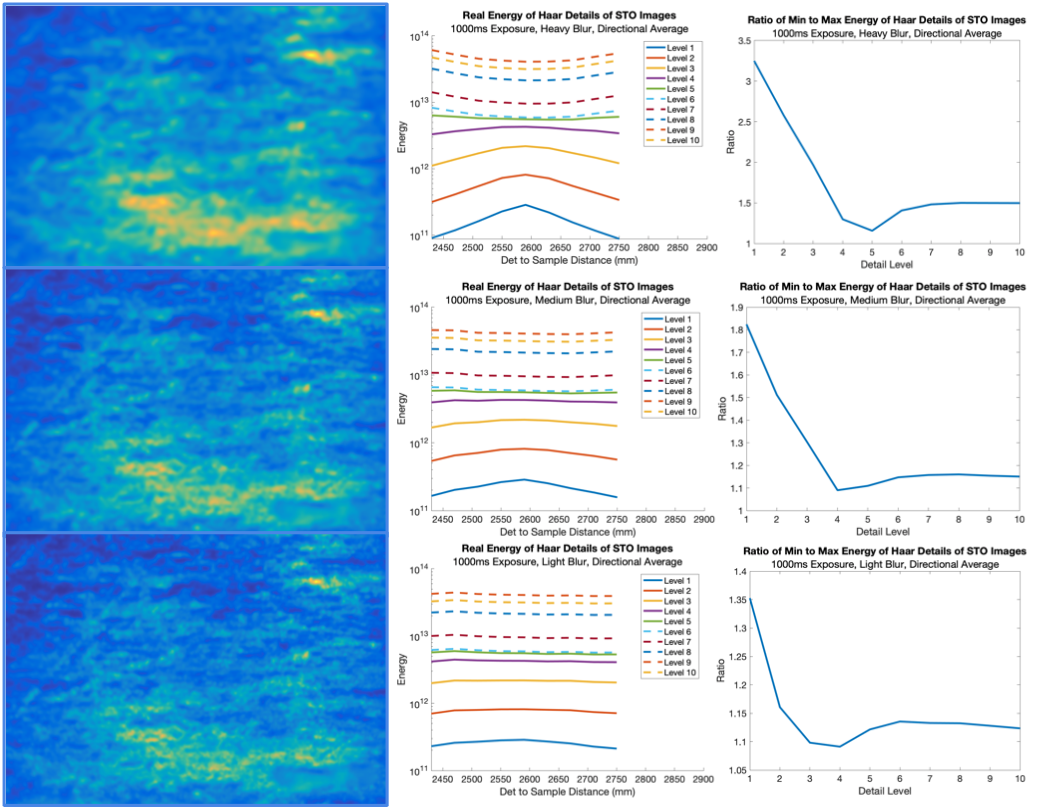}
    \caption{The left column shows contrast-normalized depictions of the most ``out-of-focus" (artificially blurred) renditions of Figure \ref{fig:imageset}C, with heavy (top), medium (middle), and light (bottom) blur. The middle column shows plots of the energy of corresponding Haar decomposition detail levels as a function of detector to sample distance, and the right column shows the ratio of minimum to maximum energies of the STO images for a given detail level. This essentially indicates the capacity for a given detail level to distinguish between the blurred and focused images. At all three levels of blur, detail level one has an energy ratio that is significantly higher than all other detail levels.}
    \label{fig:focus}
\end{figure}

While feature extraction in DFXM images motivated this analysis technique, it does not mark the full extent of its capabilities. Wavelet analysis can be used to track the sharpness of an image, thereby determining when a DFXM image is in focus. The technique utilizes the fact that sharper images have more energy in detail signals of smaller length scales compared to blurred images. A comparison of detail level energies in a set of artificially blurred DFXM images on strontium titanate (STO) is shown in Figure \ref{fig:focus}.  The center column of the figure demonstrates a clear peak in low-level detail energy and a corresponding valley in high-level detail energy for the focused image. Furthermore, the left column demonstrates that the proportional change in energy between the focused and blurred images is most pronounced in the first detail level. Thus, it was concluded that finding the image which contained the maximum energy in its first detail level would offer the most robust distinction between images in and out of focus. Visual depictions of detail levels for simulated images are shown in Appendix \ref{sec:focapp}. The Haar transform was found to provide sufficient analysis for this algorithm.
\section{Outlook}
\par  The choice of a particular wavelet in this paper was based off of qualitative and subjective measures\textemdash a quantitative method of discerning the most efficient wavelet for performing a given analysis of DFXM images of partricular system needs to be further developed. We note that this work utilized wavelets solely on single images as two-dimensional signals. Further applications and a more robust analysis may be performed if images were analyzed in sequence, forming a three-dimensional signal with two spatial dimensions and one temporal. This may allow the isolation of temporal variations to observe real-time phase transformation under dynamic physical perturbations in a spatially resolved way. One might use wavelet analyses to extract how specific domains shrink or grow under these effects, or how their orientation changes. Finally, one may utilize this algorithm to determine when a DFXM microscope is in focus, or whether a rotation axis for a sample is oriented and its location. However, the algorithm to determine image focus herein made use of data that was idealistic in nature\textemdash the algorithm should be further tested and developed on real, experimental images collected at various distances with respect to expecgted image-plane location. At next generation near diffraction-limited x-ray sources DFXM is poised to become a potent imaging modality for materials science, applied materials, and condensed matter physics research. Further exploitation of wavelet based analyses then are likely to enhance and benefit these research fields.
\newpage
\appendix
\section{Optical Image of YBCO Sample}
\begin{figure}
	\centering
	\includegraphics[angle=0,width=0.9\textwidth]{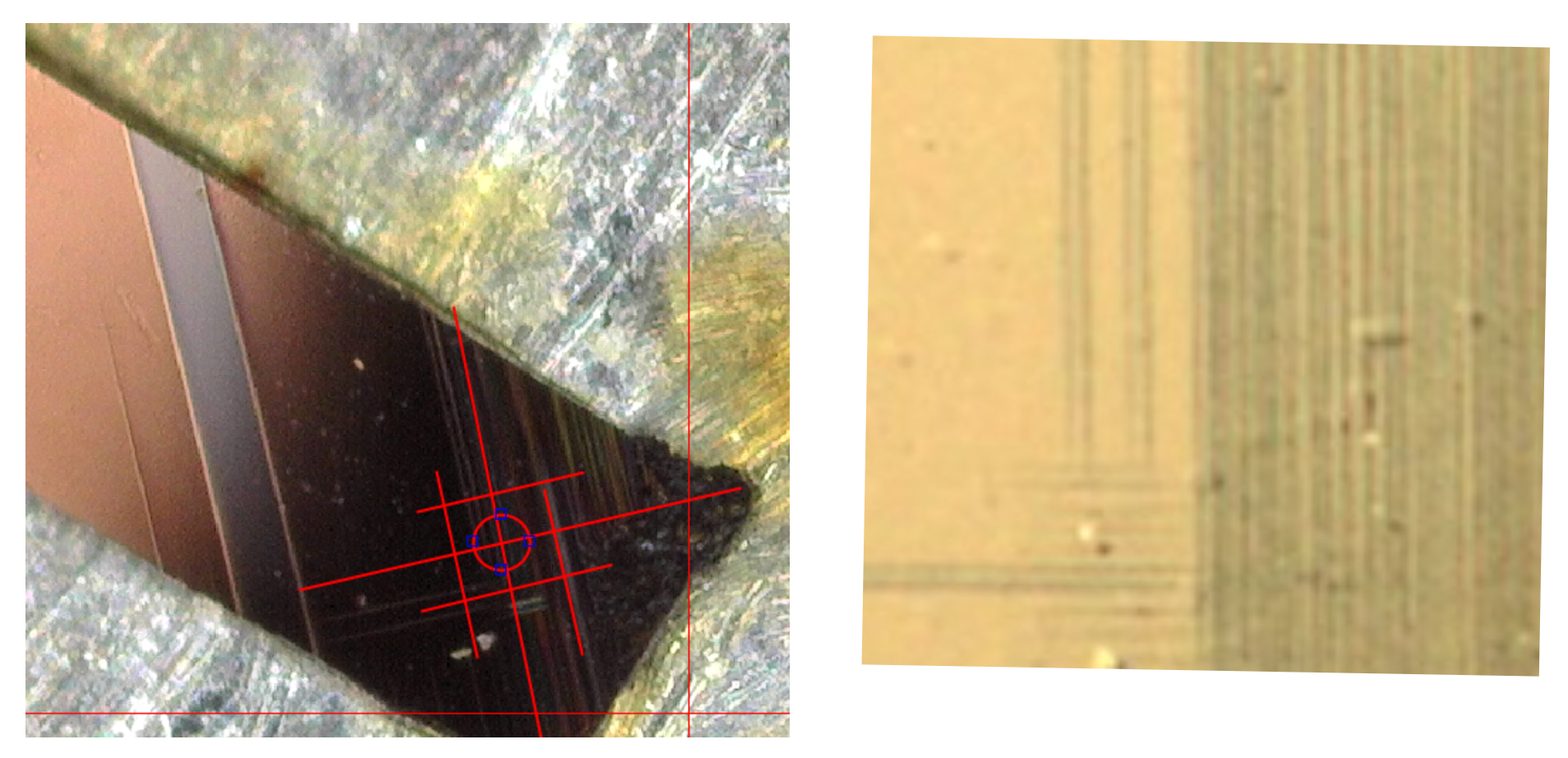}
    \caption{Left: Image of YBCO sample on Al. This sample was used to collect the data in Figure \ref{fig:imageset}A. The red square indicates the region of interest, with the crosshairs intersecting at the rotational axis. Right: Polarized microscopic imaging of region of interest.}
    \label{fig:sample}
\end{figure}
\section{More on Techniques and Methods}
The following section contains more in-depth information about the methods utilized in this report. 

\subsection{Selective Reconstruction}
\label{sec:appendixsr}
\begin{figure}
	\centering
	\centerline{\includegraphics[angle=0,width=1\textwidth]{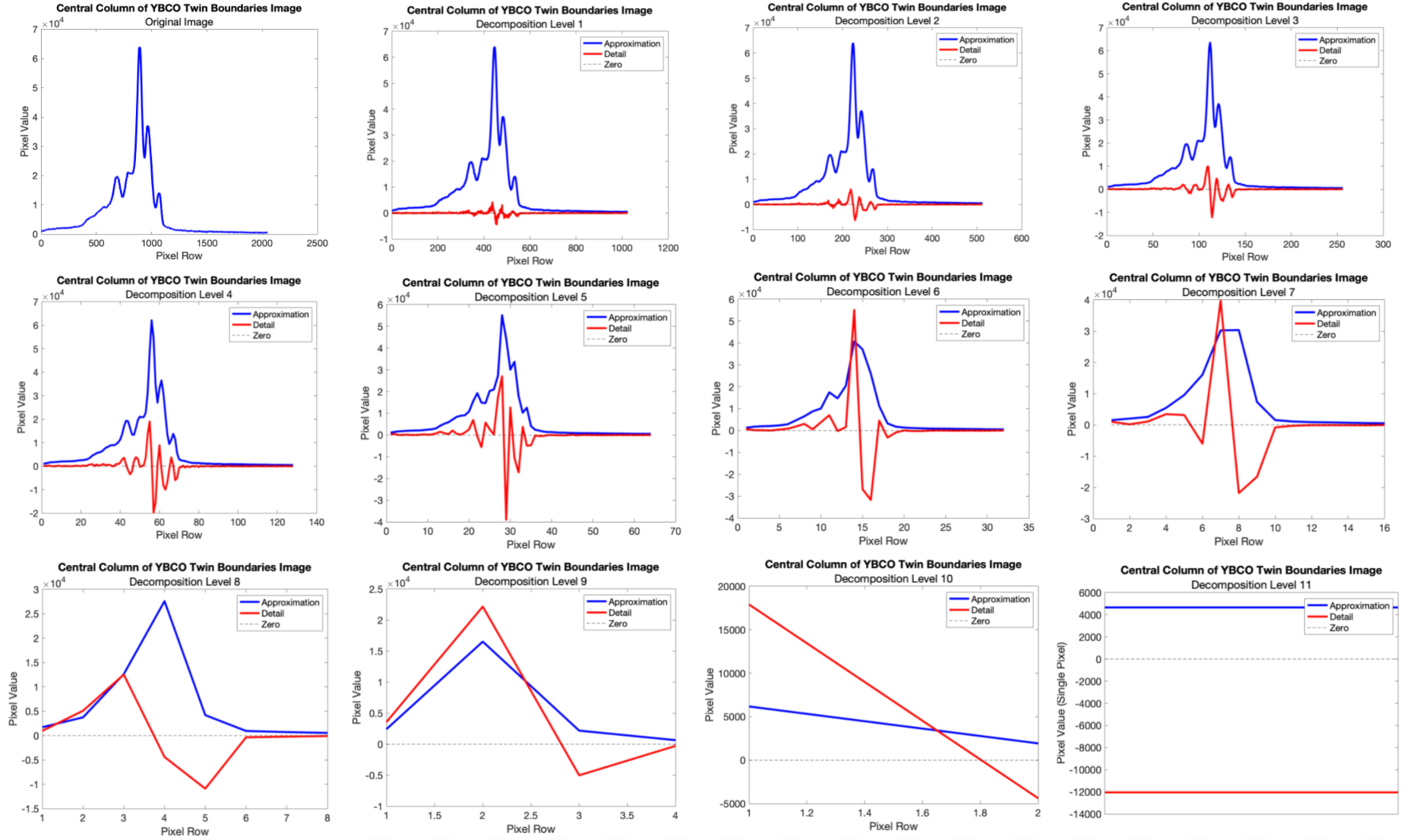}}
    \caption{Plots of central column approximation and detail signals for levels 0-11 of a Haar decomposition. In the early stages of decomposition, the small-scale features disappear first, and the amplitude of the detail signal is small compared to the approximation signal. As the decomposition level grows, the larger-scale features begin to vanish, and the detail signal begins to encapsulate smoother oscillations within the signal.}
    \label{fig:decomplevels}
\end{figure}

\begin{figure}
	\centering
	\includegraphics[angle=0,width=1\textwidth]{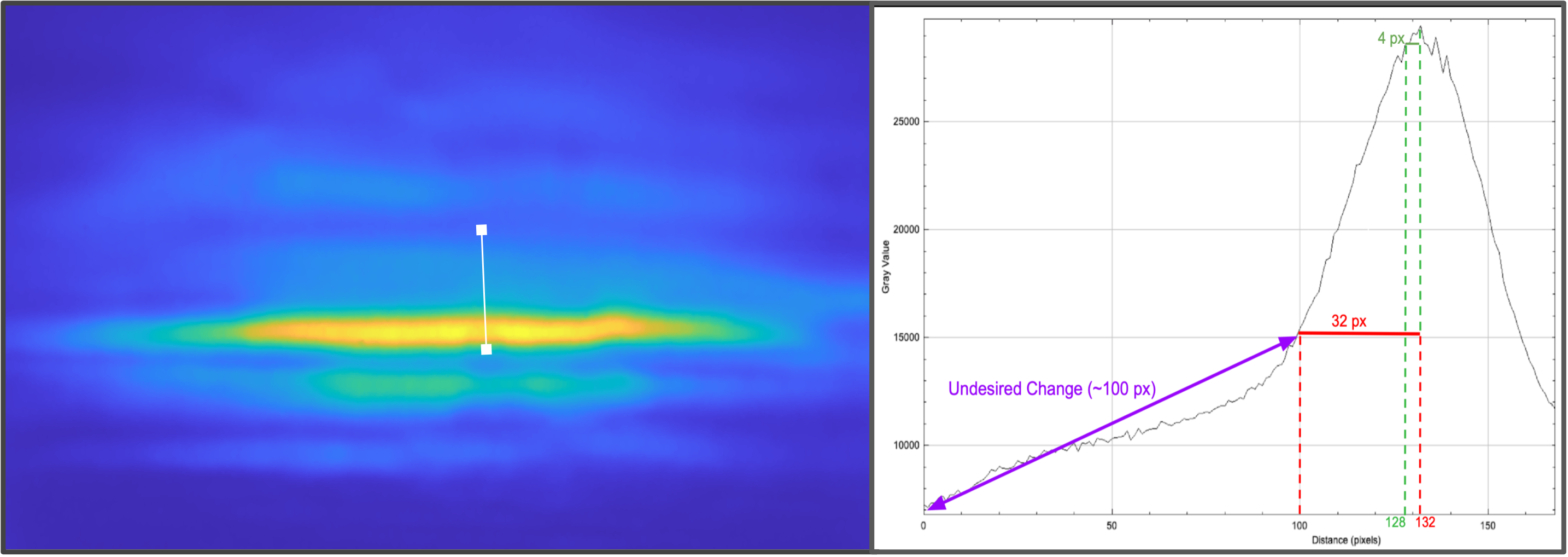}
    \caption{Line profile of Figure \ref{fig:imageset}A, with highlighted relevant changes and their corresponding length scales.}
    \label{fig:lineprof}
\end{figure}

With successive wavelet transforms, one can select which detail signals to utilize in the reconstruction, thus selectively omitting undesired features of particular frequencies in the newly reconstructed image. For example, take the signal
$$S = (1, 3, 2, 4),$$ which, under a Haar transform, has first-level approximation and detail signals $$A_1 = \sqrt{2}(2, 3)\text{ and } D_1 = \sqrt{2}(-1,-1).$$ Applying a Haar transform to $A_1$, we have $$A_2=2(2.5) \text{ and } D_2 = 2(-0.5).$$ Upon resizing $A_2$ and scaling by a factor of $\frac{1}{\sqrt{2}}$ to become $$A_2' = \sqrt{2}(2.5, 2.5),$$ reconstructing $S$ using only $A_2'$ and $D_1$ yields $$ S_{rec}=(1.5,3.5,1.5,3.5),$$ thereby removing the smooth growth of the values of $S$.\\
The decision of which detail levels to utilize in the selective reconstruction of Figure \ref{fig:imageset}A came from inspecting Figure \ref{fig:decomplevels} and preserving the levels in which desirable features were emergent. By observing the scale on which desirable changes these  occur, we find that they align with the expectations of how detail levels and the scale of changes are related, namely that a preserved detail level $\ell$ captures change on the order of $2^{\ell-1}$ pixels. Figure \ref{fig:lineprof} shows that the features in Figure \ref{fig:imageset}A have a width of approximately 64 pixels, meaning that the relevant changes between the edges and center are roughly 32 pixels in length. These are the largest desired changes, and are captured by detail level 6 as expected. The smallest desired changes are the fine details that comprise the characteristic shape of a feature, without being artifacts of noise. Figure \ref{fig:lineprof} indicates that changes that supersede the erratic jumps present within features are roughly 4 pixels in length, and are thus captured by detail level 3. 

\subsection{Feature Tracking Algorithm}
\label{sec:apptrack}
\par Given two features $F_1$ and $F_2$ with $n$ properties, indicated by $F_1(1)\dots F_1(n)$ and $F_2(1)\dots F_2(n)$, the feature distance $\Delta F_{12}$ between them is given by
\begin{equation}
    \Delta F_{12} = \sqrt{\sum_{i=1}^n \left (\frac{F_1(i)-F_2(i)}{M_i}\right )^2},
\end{equation}
where $M_i$ is a normalization parameter. For bounded properties (e.g. orientation, centroid),
$M_i$ is given by the maximum possible difference in the $i$th property; for unbounded properties (e.g. area, diameter), $M_i = \max(|F_1(i)|,|F_2(i)|)$.

\subsection{Detail Images on STO}
\label{sec:focapp}
\begin{figure}
	\centering
	\includegraphics[angle=0,width=1\textwidth]{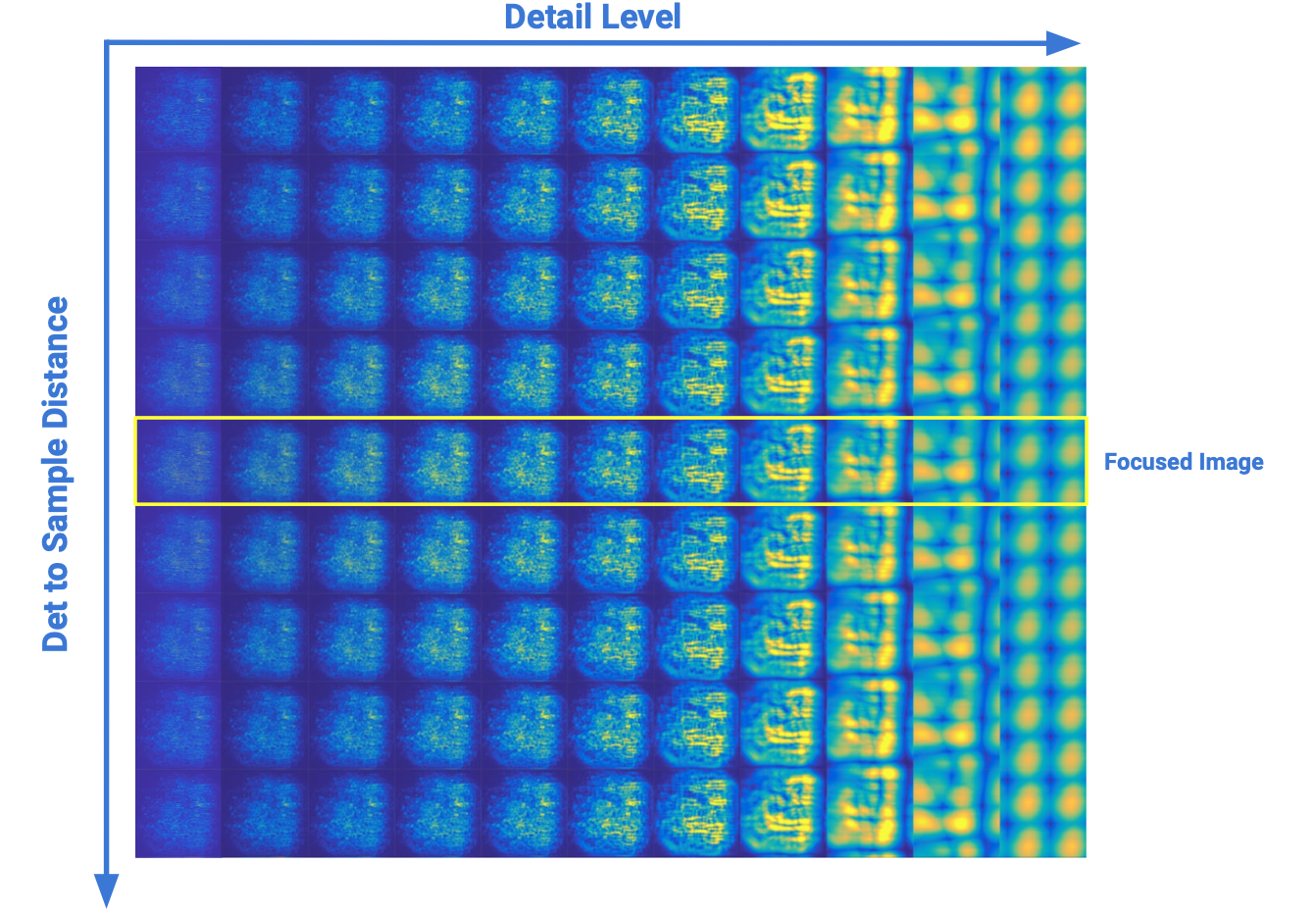}
    \caption{Grid of details of simulated STO images. The row outlined in yellow indicates the detail signals of the focused image. The images were generated with the stationary wavelet transform which differs slightly from the discrete wavelet transform in that it preserves image size by altering the downsampling and upsampling of the discrete wavelet transform. More details on the stationary transform can be found in \citeasnoun{nason1995stationary}. The images shown are the averages of the absolute values of the horizontal, vertical, and diagonal details of a given image. The periodicity of high-detail-level images is an artifact of how Matlab addresses the boundary-value problem.}
    \label{fig:dig}
\end{figure}
\ack{Acknowledgements}

We extend our gratitude to Doga Gursoy for the insights provided in our discussion on the wavelet transform and other signal analysis tools. This work was supported in part by the U.S. Department of Energy, Office of Science, Office of Workforce Development for Teachers and Scientists (WDTS) under the Science Undergraduate Laboratory Internships (SULI) program. This research used resources of the Advanced Photon Source, a U.S. Department of Energy (DOE) Office of Science user facility at Argonne National Laboratory and is based on research supported by the U.S. DOE Office of Science-Basic Energy Sciences, under Contract No. DE-AC02-06CH11357.

\referencelist[iucr]


@article{simons2018long,
  title={Long-range symmetry breaking in embedded ferroelectrics},
  author={Simons, Hugh and Haugen, Astri Bj{\o}rnetun and Jakobsen, Anders Clemen and Schmidt, S{\o}ren and St{\"o}hr, Frederik and Majkut, Marta and Detlefs, Carsten and Daniels, John E and Damjanovic, Dragan and Poulsen, Henning Friis},
  journal={Nature materials},
  volume={17},
  number={9},
  pages={814--819},
  year={2018},
  publisher={Nature Publishing Group}
}

@article{simons2015dark,
  title={Dark-field X-ray microscopy for multiscale structural characterization},
  author={Simons, Hugh and King, A and Ludwig, Wolfgang and Detlefs, C and Pantleon, Wolfgang and Schmidt, S{\o}ren and St{\"o}hr, F and Snigireva, I and Snigirev, A and Poulsen, Henning Friis},
  journal={Nature communications},
  volume={6},
  number={1},
  pages={1--6},
  year={2015},
  publisher={Nature Publishing Group}
}

@article{poulsen2018reciprocal,
  title={Reciprocal space mapping and strain scanning using X-ray diffraction microscopy},
  author={Poulsen, HF and Cook, PK and Leemreize, H and Pedersen, AF and Yildirim, C and Kutsal, M and Jakobsen, AC and Trujillo, JX and Ormstrup, J and Detlefs, C},
  journal={Journal of Applied Crystallography},
  volume={51},
  number={5},
  pages={1428--1436},
  year={2018},
  publisher={International Union of Crystallography}
}

@article{dresselhaus2021situ,
  title={In situ visualization of long-range defect interactions at the edge of melting},
  author={Dresselhaus-Marais, Leora E and Winther, Grethe and Howard, Marylesa and Gonzalez, Arnulfo and Breckling, Sean R and Yildirim, Can and Cook, Philip K and Kutsal, Mustafacan and Simons, Hugh and Detlefs, Carsten and others},
  journal={Science Advances},
  volume={7},
  number={29},
  pages={eabe8311},
  year={2021},
  publisher={American Association for the Advancement of Science}
}

@article{yildirim2020probing,
  title={Probing nanoscale structure and strain by dark-field x-ray microscopy},
  author={Yildirim, Can and Cook, Phil and Detlefs, Carsten and Simons, Hugh and Poulsen, Henning Friis},
  journal={MRS Bulletin},
  volume={45},
  number={4},
  pages={277--282},
  year={2020},
  publisher={Cambridge University Press}
}

@article{bucsek2019sub,
  title={Sub-surface measurements of the austenite microstructure in response to martensitic phase transformation},
  author={Bucsek, Ashley and Seiner, Hanu{\v{s}} and Simons, Hugh and Yildirim, Can and Cook, Phil and Chumlyakov, Yuriy and Detlefs, Carsten and Stebner, Aaron P},
  journal={Acta Materialia},
  volume={179},
  pages={273--286},
  year={2019},
  publisher={Elsevier}
}

@article{gonzalez2020methods,
  title={Methods to quantify dislocation behavior with dark-field X-ray microscopy timescans of single-crystal aluminum},
  author={Gonzalez, Arnulfo and Howard, Marylesa and Breckling, Sean and Dresselhaus-Marais, Leora E},
  journal={arXiv preprint arXiv:2008.04972},
  year={2020}
}

@article{flannery1987three,
  title={Three-dimensional X-ray microtomography},
  author={Flannery, Brian P and Deckman, Harry W and Roberge, Wayne G and D'Amico, Kevin L},
  journal={Science},
  volume={237},
  number={4821},
  pages={1439--1444},
  year={1987},
  publisher={American Association for the Advancement of Science}
}

@article{heil1989continuous,
  title={Continuous and discrete wavelet transforms},
  author={Heil, Christopher E and Walnut, David F},
  journal={SIAM review},
  volume={31},
  number={4},
  pages={628--666},
  year={1989},
  publisher={SIAM}
}

@article{shensa1992discrete,
  title={The discrete wavelet transform: wedding the a trous and Mallat algorithms},
  author={Shensa, Mark J and others},
  journal={IEEE Transactions on signal processing},
  volume={40},
  number={10},
  pages={2464--2482},
  year={1992}
}

@article{ludwig2001three,
  title={Three-dimensional imaging of crystal defects bytopo-tomography'},
  author={Ludwig, W and Cloetens, P and H{\"a}rtwig, J and Baruchel, J and Hamelin, B and Bastie, P},
  journal={Journal of applied crystallography},
  volume={34},
  number={5},
  pages={602--607},
  year={2001},
  publisher={International Union of Crystallography}
}

@article{karpov2017three,
  title={Three-dimensional imaging of vortex structure in a ferroelectric nanoparticle driven by an electric field},
  author={Karpov, D and Liu, Z and Rolo, T and Harder, R and Balachandran, PV and Xue, D and Lookman, T and Fohtung, E},
  journal={Nature communications},
  volume={8},
  number={1},
  pages={1--8},
  year={2017},
  publisher={Nature Publishing Group}
}

@article{dupraz20173d,
  title={3D imaging of a dislocation loop at the onset of plasticity in an indented nanocrystal},
  author={Dupraz, Maxime and Beutier, Guillaume and Cornelius, Thomas W and Parry, Guillaume and Ren, Zhe and Labat, St{\'e}phane and Richard, M-I and Chahine, Gilbert A and Kovalenko, Oleg and De Boissieu, Marc and others},
  journal={Nano letters},
  volume={17},
  number={11},
  pages={6696--6701},
  year={2017},
  publisher={ACS Publications}
}

@book{walker2008primer,
  title={A primer on wavelets and their scientific applications},
  author={Walker, James S},
  year={2008},
  publisher={Chapman and hall/CRC}
}

@article{brennan2022ana,
  title={Analytical methods for superresolution dislocation identification in dark-field X-ray microscopy},
  author={Brennan, Michael C. and Howard, Marylesa and Marzouk, Youssef and Dresselhaus-Marais, Leora E.},
  journal={Journal of Materials Science},
  volume={57},
  number={31},
  pages={14890--14904},
  year={2022},
  publisher={Springer Nature }
}

@techreport{tamraz,
    title = {Intelligent optical imaging to steer and enhance high-resolution dark-field x-ray microscopy at the advanced photon source},
    author = {Tamraz, Rokia },
    year={2021},
    institution={Advanced Photon Source (APS)(United States)}
}

@incollection{nason1995stationary,
  title={The stationary wavelet transform and some statistical applications},
  author={Nason, Guy P and Silverman, Bernard W},
  booktitle={Wavelets and statistics},
  pages={281--299},
  year={1995},
  publisher={Springer}
}

@article{qiao2020large,
  title={A large field-of-view high-resolution hard x-ray microscope using polymer optics},
  author={Qiao, Zhi and Shi, Xianbo and Kenesei, Peter and Last, Arndt and Assoufid, Lahsen and Islam, Zahir},
  journal={Review of Scientific Instruments},
  volume={91},
  number={11},
  pages={113703},
  year={2020},
  publisher={AIP Publishing LLC}
}

@article{poudyal2022pump,
  title={Pump-probe Dark-field X-ray Microscopy},
  author={Poudyal, Ishwor and Qiao, Zhi and Armstrong, Michael R and Islam, Zahir},
  journal={arXiv preprint arXiv:2210.06243},
  year={2022}
}

@techreport{crabtree2012quanta,
  title={From quanta to the continuum: opportunities for mesoscale science},
  author={Crabtree, George and Sarrao, John and Alivisatos, Paul and Barletta, William and Bates, Frank and Brown, Gordon and French, Roger and Greene, Laura and Hemminger, John and Kastner, Marc and others},
  year={2012},
  institution={USDOE Office of Science (SC)(United States)}
}
\end{document}